\documentclass[aps,prl,twocolumn,superscriptaddress,floatfix,nofootinbib,showpacs,amsmath,amssymb,altaffilletter]{revtex4-1}

\usepackage{graphicx}% Include figure files
\usepackage{dcolumn}% Align table columns on decimal point
\usepackage{bm}% Bold math
\usepackage{color}% Colored text
\usepackage{verbatim}% Allows long comments
\usepackage{paralist}% Compact environments
\usepackage{lineno}% Line numbers
\usepackage{subfigure}

\newcommand{\ber}{\mbox{$^{7}$Be}}
\newcommand{\bor}{\mbox{$^{8}$B}}
\newcommand{\cele}{\mbox{$^{11}$C}}

\newcommand{\kr}{\mbox{$^{85}$Kr}}

\newcommand{\rbm}{\mbox{$^{\rm 85m}$Rb}}
\newcommand{\bite}{\mbox{$^{210}$Bi}}
\newcommand{\pote}{\mbox{$^{210}$Po}}
\newcommand{\pbfo}{\mbox{$^{214}$Pb}}
\newcommand{\Bipo}{\mbox{$^{214}$Bi-$^{214}$Po}}

\newcommand{\po}{\mbox{$^{218}$Po}}
\newcommand{\radon}{\mbox{$^{222}$Rn}}

\newcommand{\Pee}{\mbox{$P_{ee}$}}

\newcommand{\fbe}{\mbox{$f_{\rm ^7Be}$}}

\newcommand{\fCNO}{\mbox{$f_{\rm CNO}$}}

\newcommand{\cpd}{\mbox{counts/(day$\cdot$100\,ton)}}

\begin{document}
\title{Precision measurement of the \ber\ solar neutrino interaction rate in Borexino}

\newcommand{\APC}{Laboratoire AstroParticule et Cosmologie, 75231 Paris cedex 13, France}
\newcommand{\Budapest}{KFKI-RMKI, 1121 Budapest, Hungary}
\newcommand{\Dubna}{Joint Institute for Nuclear Research, 141980 Dubna, Russia}
\newcommand{\Genova}{Dipartimento di Fisica, Universit\`a e INFN, Genova 16146, Italy}
\newcommand{\Hamburg}{Institut f\"ur Experimentalphysik, Universit\"at, 22761 Hamburg, Germany}
\newcommand{\Heidelberg}{Max-Planck-Institut f\"ur Kernphysik, 69029 Heidelberg, Germany}
\newcommand{\Kiev}{Kiev Institute for Nuclear Research, 06380 Kiev, Ukraine}
\newcommand{\Krakow}{M.~Smoluchowski Institute of Physics, Jagiellonian University, 30059 Krakow, Poland}
\newcommand{\Kurchatov}{NRC Kurchatov Institute, 123182 Moscow, Russia}
\newcommand{\LNGS}{INFN Laboratori Nazionali del Gran Sasso, SS 17 bis Km 18+910, 67010 Assergi (AQ), Italy}
\newcommand{\Milano}{Dipartimento di Fisica, Universit\`a degli Studi e INFN, 20133 Milano, Italy}
\newcommand{\Munich}{Physik Department, Technische Universit\"at Muenchen, 85747 Garching, Germany}
\newcommand{\Pavia}{INFN, Pavia 27100, Italy}
\newcommand{\Perugia}{Dipartimento di Chimica, Universit\`a e INFN, 06123 Perugia, Italy}
\newcommand{\Peters}{St. Petersburg Nuclear Physics Institute, 188350 Gatchina, Russia}
\newcommand{\Princeton}{Physics Department, Princeton University, Princeton, NJ 08544, USA}
\newcommand{\PrincetonChemEng}{Chemical Engineering Department, Princeton University, Princeton, NJ 08544, USA}
\newcommand{\Queens}{Physics Department, Queen's University, Kingston ON K7L 3N6, Canada}
\newcommand{\UMass}{Physics Department, University of Massachusetts, Amherst, MA 01003, USA}
\newcommand{\Virginia}{Physics Department, Virginia Polytechnic Institute and State University, Blacksburg, VA 24061, USA}
\newcommand{\Valencia}{Instituto de F\`isica Corpuscular, CSIC-UVEG, Valencia 46071, Spain}

\author{G.~Bellini}\affiliation{\Milano}
\author{J.~Benziger}\affiliation{\PrincetonChemEng}
\author{D.~Bick}\affiliation{\Hamburg}
\author{S.~Bonetti}\affiliation{\Milano}
\author{G.~Bonfini}\affiliation{\LNGS}
\author{M.~Buizza~Avanzini}\affiliation{\Milano}
\author{B.~Caccianiga}\affiliation{\Milano}
\author{L.~Cadonati}\affiliation{\UMass}
\author{F.~Calaprice}\affiliation{\Princeton}
\author{C.~Carraro}\affiliation{\Genova}
\author{P.~Cavalcante}\affiliation{\LNGS}
\author{A.~Chavarria}\affiliation{\Princeton}
\author{D.~D'Angelo}\affiliation{\Milano}
\author{S.~Davini}\affiliation{\Genova}
\author{A.~Derbin}\affiliation{\Peters}
\author{A.~Etenko}\affiliation{\Kurchatov}\affiliation{\LNGS}
\author{K.~Fomenko}\affiliation{\Dubna}\affiliation{\LNGS}
\author{D.~Franco}\affiliation{\APC}
\author{C.~Galbiati}\affiliation{\Princeton}
\author{S.~Gazzana}\affiliation{\LNGS}
\author{C.~Ghiano}\affiliation{\LNGS}
\author{M.~Giammarchi}\affiliation{\Milano}
\author{M.~Goeger-Neff}\affiliation{\Munich}
\author{A.~Goretti}\affiliation{\Princeton}
\author{L.~Grandi}\affiliation{\Princeton}
\author{E.~Guardincerri}\affiliation{\Genova}
\author{S.~Hardy}\affiliation{\Virginia}
\author{Aldo~Ianni}\affiliation{\LNGS}
\author{Andrea~Ianni}\affiliation{\Princeton}
\author{V.~Kobychev}\affiliation{\Kiev}
\author{D.~Korablev}\affiliation{\Dubna}
\author{G.~Korga}\affiliation{\LNGS}
\author{Y.~Koshio}\affiliation{\LNGS}
\author{D.~Kryn}\affiliation{\APC}
\author{M.~Laubenstein}\affiliation{\LNGS}
\author{T.~Lewke}\affiliation{\Munich}
\author{E.~Litvinovich}\affiliation{\Kurchatov}
\author{B.~Loer}\affiliation{\Princeton}
\author{F.~Lombardi}\affiliation{\LNGS}
\author{P.~Lombardi}\affiliation{\Milano}
\author{L.~Ludhova}\affiliation{\Milano}
\author{I.~Machulin}\affiliation{\Kurchatov}
\author{S.~Manecki}\affiliation{\Virginia}
\author{W.~Maneschg}\affiliation{\Heidelberg}
\author{G.~Manuzio}\affiliation{\Genova}
\author{Q.~Meindl}\affiliation{\Munich}
\author{E.~Meroni}\affiliation{\Milano}
\author{L.~Miramonti}\affiliation{\Milano}
\author{M.~Misiaszek}\affiliation{\Krakow}\affiliation{\LNGS}
\author{D.~Montanari}\affiliation{\LNGS}\affiliation{\Princeton}
\author{P.~Mosteiro}\affiliation{\Princeton}
\author{V.~Muratova}\affiliation{\Peters}
\author{L.~Oberauer}\affiliation{\Munich}
\author{M.~Obolensky}\affiliation{\APC}
\author{F.~Ortica}\affiliation{\Perugia}
\author{M.~Pallavicini}\affiliation{\Genova}
\author{L.~Papp}\affiliation{\LNGS}
\author{C.~Pe\~na-Garay}\affiliation{\Valencia}
\author{L.~Perasso}\affiliation{\Milano}
\author{S.~Perasso}\affiliation{\Genova}
\author{A.~Pocar}\affiliation{\UMass}
\author{R.S.~Raghavan}\affiliation{\Virginia}
\author{G.~Ranucci}\affiliation{\Milano}
\author{A.~Razeto}\affiliation{\LNGS}
\author{A.~Re}\affiliation{\Milano}
\author{A.~Romani}\affiliation{\Perugia}
\author{A.~Sabelnikov}\affiliation{\Kurchatov}
\author{R.~Saldanha}\affiliation{\Princeton}
\author{C.~Salvo}\affiliation{\Genova}
\author{S.~Sch\"onert}\affiliation{\Munich}
\author{H.~Simgen}\affiliation{\Heidelberg}
\author{M.~Skorokhvatov}\affiliation{\Kurchatov}
\author{O.~Smirnov}\affiliation{\Dubna}
\author{A.~Sotnikov}\affiliation{\Dubna}
\author{S.~Sukhotin}\affiliation{\Kurchatov}
\author{Y.~Suvorov}\affiliation{\LNGS}
\author{R.~Tartaglia}\affiliation{\LNGS}
\author{G.~Testera}\affiliation{\Genova}
\author{D.~Vignaud}\affiliation{\APC}
\author{R.B.~Vogelaar}\affiliation{\Virginia}
\author{F.~von~Feilitzsch}\affiliation{\Munich}
\author{J.~Winter}\affiliation{\Munich}
\author{M.~Wojcik}\affiliation{\Krakow}
\author{A.~Wright}\affiliation{\Princeton}
\author{M.~Wurm}\affiliation{\Munich}
\author{J.~Xu}\affiliation{\Princeton}
\author{O.~Zaimidoroga}\affiliation{\Dubna}
\author{S.~Zavatarelli}\affiliation{\Genova}
\author{G.~Zuzel}\affiliation{\Heidelberg}

\collaboration{Borexino Collaboration}
\noaffiliation

\date{\today}

\begin{abstract}
The rate of neutrino-electron elastic
scattering interactions from 862~keV \ber\ solar neutrinos in Borexino
is determined to be
46.0$\pm$1.5$({\rm stat})$$^{+1.5}_{-1.6}({\rm syst})$\,\cpd. This
corresponds to a $\nu_e$-equivalent \ber\ solar neutrino flux of  (3.10$\pm$0.15)$\times$$10^9$\,cm$^{-2}$s$^{-1}$ and, under the
assumption of $\nu_e$ transition to other active neutrino flavours,
yields an electron neutrino survival probability of
0.51$\pm$0.07 at 862~keV. The no flavor change hypothesis is ruled out
at 5.0\,$\sigma$. A global solar neutrino analysis
with free fluxes determines
$\Phi_{pp}$=6.06$^{+0.02}_{-0.06}$$\times$$10^{10}$\,cm$^{-2}$s$^{-1}$
and
$\Phi_{CNO}$$<$1.3$\times$$10^9$\,cm$^{-2}$s$^{-1}$~(95\%~C.L.). These
results significantly improve the precision with which the MSW-LMA neutrino
oscillation model is experimentally tested at low energy.  
\end{abstract}


\maketitle

In the past 40 years, solar neutrino
experiments~\cite{bib:rchem-cl,bib:rchem-ga,bib:kamiokande,bib:sno}
have revealed important information about the
Sun~\cite{bib:bahcallSSM} and have shown that solar neutrinos undergo flavour transitions that are well described by
Mikheyev-Smirnov-Wolfenstein Large Mixing Angle (``MSW-LMA'') type flavour
oscillations~\cite{bib:msw}. Reactor antineutrino
measurements~\cite{bib:kamland} also support this model. The MSW model
predicts a transition in the solar $\nu_e$ survival probability
(``\Pee'') at neutrino
energies of about 1-4\,MeV. This transition is currently poorly
tested. Therefore, in order to test
MSW-LMA more thoroughly, to probe other proposed neutrino
oscillation scenarios~\cite{bib:nonstandard}, and to further improve
our understanding of the Sun, it is important
that experimental measurements of the low energy solar
neutrino fluxes be improved~\cite{bib:bahcallLow}. At 862\,keV, the abundant,
mono-energetic, \ber\ solar neutrinos can provide a precise probe of the
survival probability in this interesting region.  In addition, a
precise determination of the \ber\ flux combined with existing
results from radiochemical
experiments~\cite{bib:rchem-cl,bib:rchem-ga} yields improved
constraints on the $pp$ and CNO solar neutrino fluxes.

The Borexino experiment at Gran Sasso~\cite{bib:bxdetectorpaper} detects neutrinos through the neutrino-electron elastic
scattering interaction on a $\sim$278 metric ton liquid scintillator target. The
low energy backgrounds in the detector have been suppressed to
unprecedented levels~\cite{bib:bxfirstresults}, making Borexino
the first experiment capable of making spectrally resolved
measurements of solar neutrinos at energies below 1\,MeV. We
have previously reported a direct
measurement of the \ber\ solar neutrino flux with combined statistical and
systematic errors of
10\%~\cite{bib:bxsecondresults}. Following a campaign of detector calibrations and a 4-fold
increase in solar neutrino exposure, we present here a new \ber\ neutrino
flux measurement with a total uncertainty less than 5\%. For the first time, the
experimental uncertainty is smaller than the uncertainty in the Standard Solar Model
(``SSM'') prediction of the
\ber\ neutrino flux~\cite{bib:ssm2011}\footnote{Throughout this Letter
  we use the high metallicity SSM predictions from the ``GS98'' column
  in Table 2 of~\cite{bib:ssm2011} as our reference SSM. For comparison, the
  \ber\ neutrino flux predicted by the low metallicity model (also
  given in~\cite{bib:ssm2011}) is 8.8\% lower than the high
  metallicity prediction.}.

The new result is based on the analysis of 740.7 live
days (after cuts) of data which were recorded in the period from May 16, 2007
to May 8, 2010, and which correspond to a 153.6\,ton$\cdot$yr
fiducial exposure. 

The experimental signature of \ber\ neutrino interactions in
Borexino is a Compton-like shoulder at $\sim$660\,keV. Fits to the
spectrum of observed event energies are used to distinguish between
this neutrino scattering feature and backgrounds from radioactive decays~\cite{bib:bxsecondresults}. Two
independent fit methods
were used, one which is Monte Carlo based and one which uses an
analytic description of the detector response. In both methods, the
weights for the \ber\ neutrino signal and the main radioactive
background components (\kr, $^{210}$Po, \bite, and \cele) were left as free parameters in
the fit,
while the contributions of the {\it pp}, {\it pep}, CNO, and \bor\
solar neutrinos were fixed
to the SSM-predicted rates assuming MSW neutrino oscillations with
$\tan^2{\theta_{12}}$=0.47$^{+0.05}_{-0.04}$ and $\Delta
m_{12}^2$=(7.6$\pm$0.2)$\times$$10^{-5}$\,eV$^2$~\cite{bib:pdg2010}. The
impact of fixing these fluxes was evaluated and
included as a systematic uncertainty. The rates of \radon, \po, and \pbfo\ surviving
the cuts were fixed using the measured rate of \Bipo\ delayed
coincidence events. The Monte Carlo method also includes external $\gamma$-ray
background, which makes it possible to extend the fit range in this
method to higher energies. The energy scale and resolution were floated in the
analytic fits, while the Monte Carlo approach automatically
incorporates the simulated energy response of the detector.

\begin{figure}[!t]
\subfigure{}{\includegraphics[width=0.45\textwidth]{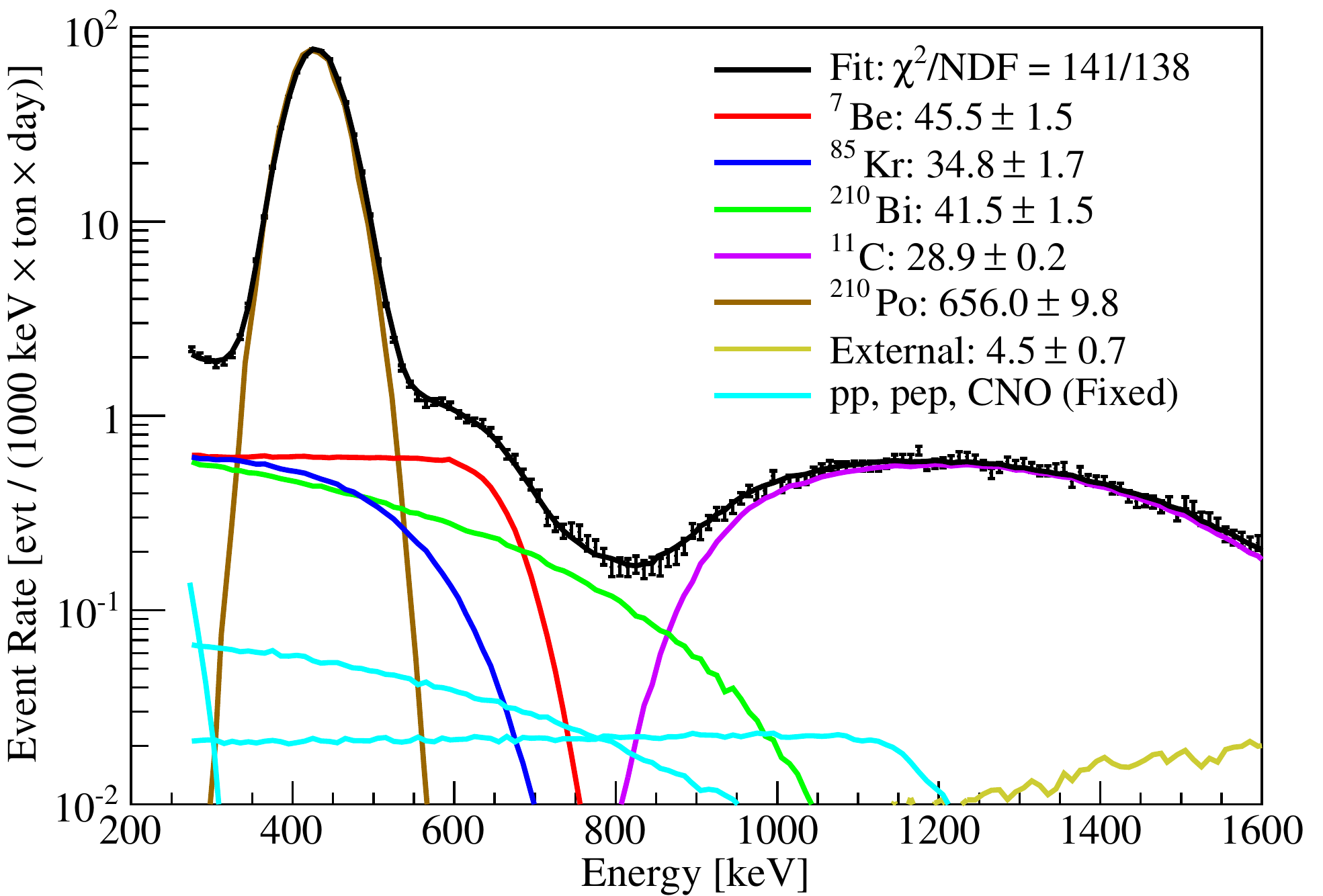}}
\subfigure{}{\includegraphics[width=0.45\textwidth]{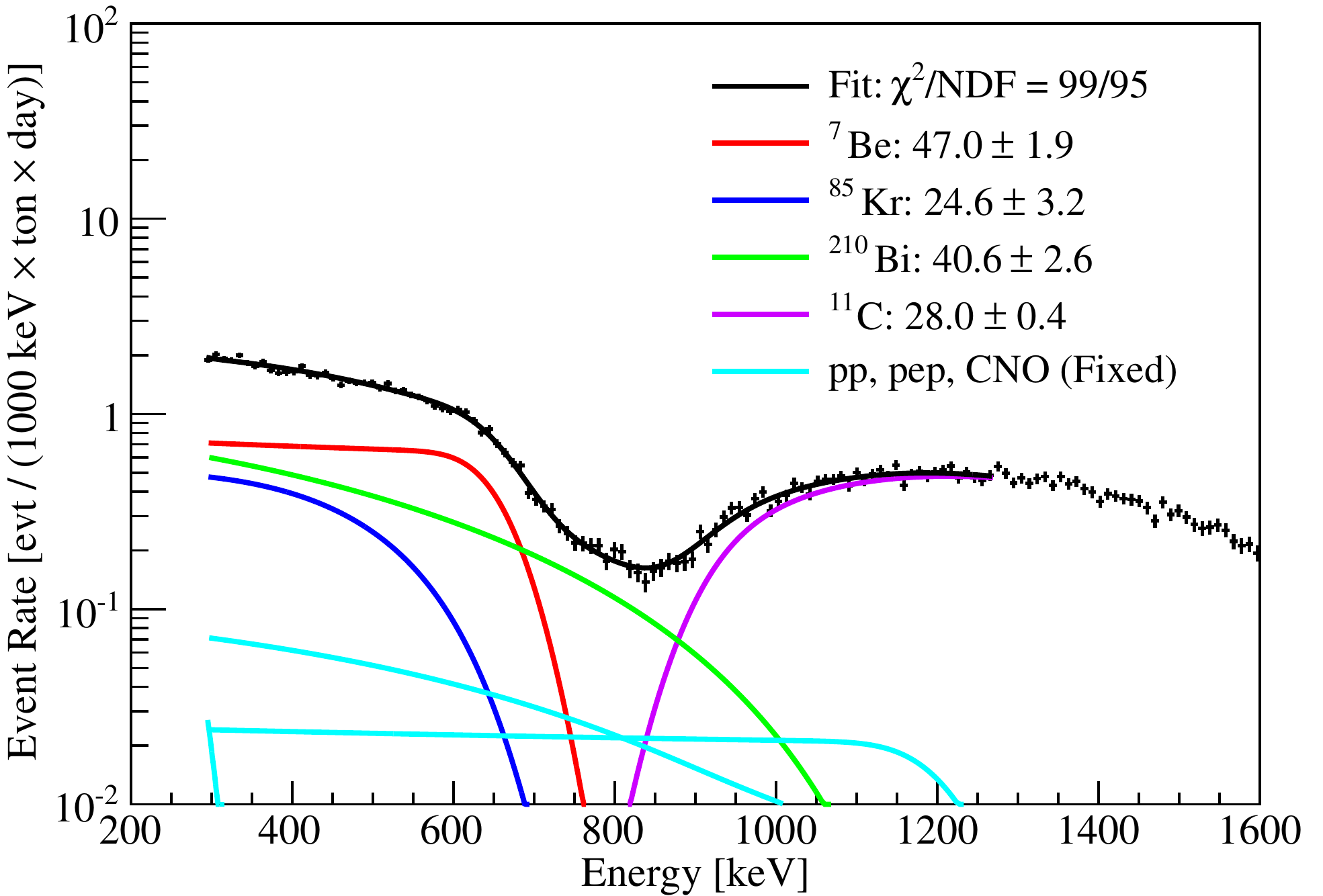}}
\caption{Two example fitted spectra; the fit results in the legends
  have units [\cpd]. Top: A Monte Carlo based fit over
  the energy region 270--1600~keV to a spectrum from which some, but not
  all, of the $\alpha$ events have been removed using a
  PSA cut, and in which the event energies were estimated using the number of
  photons detected by the PMT array. Bottom: An analytic fit over the
  290--1270~keV energy region to a spectrum obtained with statistical
  $\alpha$ subtraction and in which the event energies were
  estimated using the total charge collected by the PMT array. In all
  cases the fitted event rates refer to the total rate of each
  species, independent of the fit energy window.}
\label{fig:fit}
\end{figure}

The stability of each fit method was studied by
repeating the fits with slightly varied fit characteristics (e.g. fit range and
histogram binning) and different methods of data
preparation. The latter included changing the method used to estimate
the event energies, and varying the pulse shape
analysis (``PSA'') technique~\cite{bib:gatti} used to remove \pote\ and other
$\alpha$ events between a highly efficient statistical
subtraction method~\cite{bib:bxsecondresults} and a cut-based
technique which removes a fraction of the $\alpha$ events with a very
small loss of
$\beta$ events. The example spectra shown in Fig.~\ref{fig:fit} illustrate
the stability of our fit procedure; the \bor\ neutrino and 
\pbfo, \radon, and \po\ background spectra are small on the
scale of the plots and are not shown. The results of these and
other fits using different permutations of the fit characteristics and data
preparation techniques described above were averaged to obtain the
central values reported in Table~\ref{tab:fit-results}; the spread
between the results is included in the systematic uncertainty.

\begin{table}[!b]
\begin{center}
\caption{Average Fit Results [\cpd].}
\begin{tabular}{lc}
\hline\hline
\ber\			&46.0$\pm$1.5$({\rm stat})$$^{+1.5}_{-1.6}({\rm syst})$ \\
\kr\			&31.2$\pm$1.7$({\rm stat})$$\pm$4.7$({\rm syst})$ \\
\bite			&41.0$\pm$1.5$({\rm stat})$$\pm$2.3$({\rm syst})$ \\
\cele\		&28.5$\pm$0.2$({\rm stat})$$\pm$0.7$({\rm syst})$ \\
\hline\hline
\end{tabular}
\label{tab:fit-results}
\end{center}
\end{table}

The main systematic uncertainties in our measurement of the
\ber\ interaction rate are listed in Table~\ref{tab:systerr}.  The dominant contributions come from the
determination of the fiducial volume, our understanding of the
detector energy response, and the variation between the results of the
different fit procedures. We note the significant decrease in the
uncertainties associated with the
detector energy response and the definition of the fiducial volume, by factors of 4.6 and 2.2 respectively, relative
to~\cite{bib:bxsecondresults}. This improvement was made possible by several campaigns of detector calibration.

\begin{table}[!t]
\begin{center}
\caption{\ber\ Systematic Uncertainties [\%].}
\begin{tabular}{lc}
\hline\hline
Source					&[\%] \\
\hline
Trigger efficiency and stability	&$<$0.1 \\
Live time				&0.04 \\
Scintillator density			&0.05 \\
Sacrifice of cuts			&0.1 \\
Fiducial volume 		&$^{+0.5}_{-1.3}$ \\
Fit methods				&2.0 \\
Energy response			&2.7 \\
\hline
Total Systematic Error		&$^{+3.4}_{-3.6}$ \\
\hline\hline
\end{tabular}
\label{tab:systerr}
\end{center}
\end{table}

During the first 1.6 years of Borexino operation, the energy and
position reconstruction algorithms were tuned, and their performances
estimated, using
intrinsic activities such as $^{14}$C, $^{210}$Po, and
$^{11}$C. The first deployed source calibrations were carried out in 4
campaigns between October 2008 and July 2009. Encapsulated radioactive sources, including $^{57}$Co, $^{139}$Ce, $^{203}$Hg, $^{85}$Sr, $^{54}$Mn, $^{65}$Zn,
$^{40}$K, $^{60}$Co, and $^{222}$Rn, were placed inside the scintillator volume using a rod-based
source deployment system. Using seven CCD cameras mounted within the
PMT array, the positions of the sources could be
determined with a precision better than 2\,cm.  In aggregate, more than 35 live-days of calibration data
were recorded with sources at more than 250 positions within the
scintillator volume.

The systematic uncertainty in the
definition of the fiducial mass was evaluated
by comparing the position of the source as measured by the CCD
camera system and as reconstructed using the PMT array for a number of runs with sources deployed near the boundary of the
fiducial volume. The $\gamma$-ray sources were also used to validate
and improve both the Monte Carlo (the simulated and observed
optical responses agree at the 1.5\% level within the fiducial volume) and the
detector energy response function used in the analytic fitting procedure. The uncertainties in these
tunings and in our understanding of the calibration data were included
in the energy scale systematic. A dominant contribution to the latter came from the uncertainty in the
scintillator quenching model used to extrapolate the
detector optical response from the $\gamma$-ray events used for energy calibration
to the single electron events which comprise the majority of the
data. 

It may be noted in Table~\ref{tab:fit-results} that the \kr\ rate has a larger systematic uncertainty than the
other fitted rates. This is due to a larger variation in the \kr\ rate
between the different fit procedures. We typically
obtain larger \kr\ values when  fewer $\alpha$
events are removed from the spectrum before fitting and when the fit range is extended to encompass lower
energy events.  However, as the
\ber\ rate is mostly constrained by the spectral region between 550
and 750\,keV, the variation in the \kr\ rate reflects only weakly
in the \ber\ rate. We note that the fitted \kr\ rate is
consistent with an independent measurement of 30.4$\pm$5.3(stat)$\pm$1.3(syst)\,\cpd\
for the \kr\ activity obtained using \kr-\rbm\ delayed coincidences.

\begin{figure}[!t]
\includegraphics[width=0.45\textwidth]{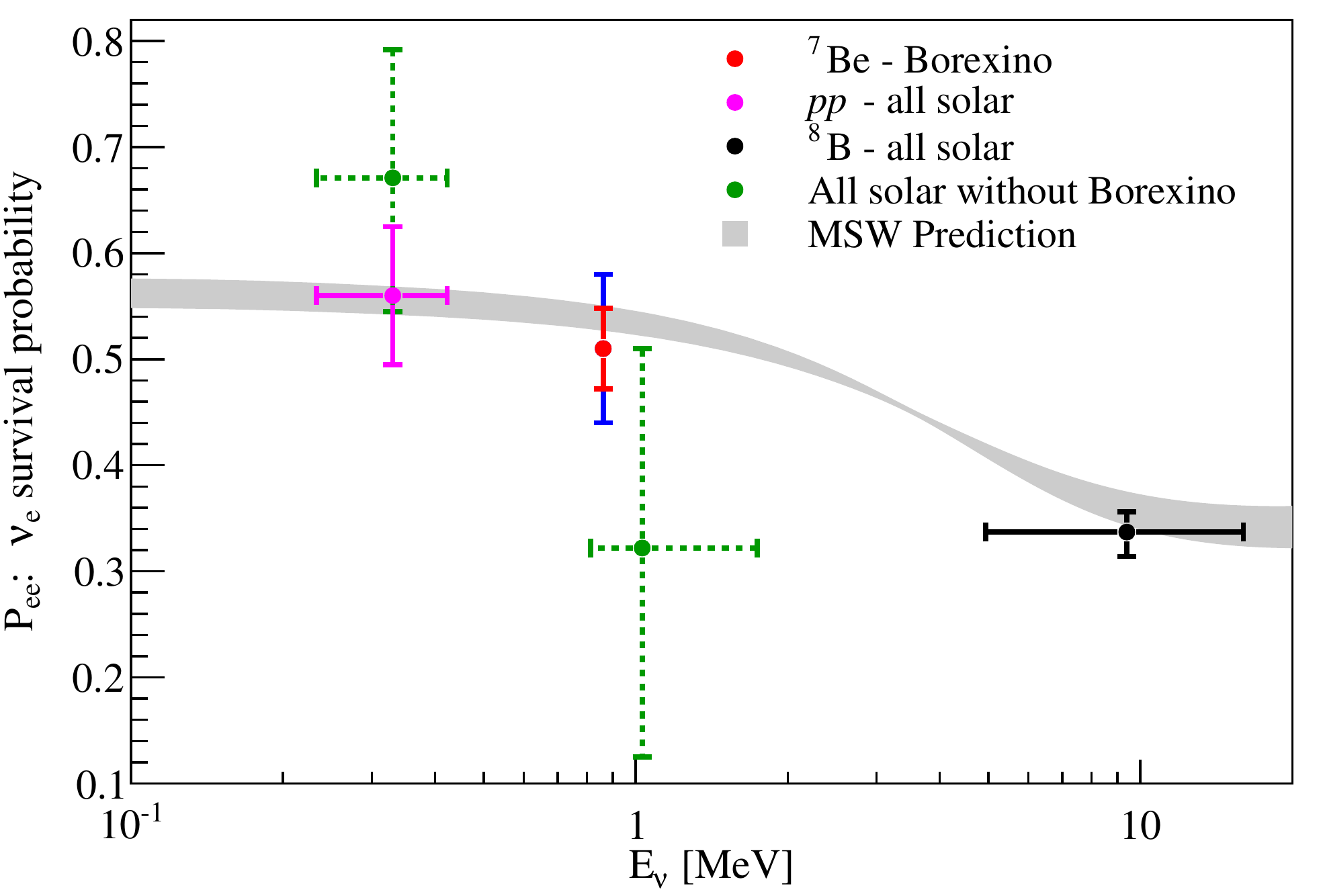}
\caption{The global experimental constraints on the low energy solar
  \Pee. For the \ber\ point, which shows the current result, the
  inner (red) error bars show the experimental uncertainty, while the outer
  (blue) error bars show the total (experimental + SSM)
  uncertainty. The remaining points were obtained
  following the procedure in~\cite{bib:barger}, wherein the survival
  probabilities of the low energy ($pp$), medium energy, and high
  energy ($^8$B) solar neutrinos are obtained, with minimal model
  dependence, from a combined analysis of the results of all solar neutrino
  experiments. To illustrate Borexino's effect on the
  low energy \Pee\ measurements, the green (dashed) points are calculated
  without using the Borexino data. The MSW-LMA
  prediction is also shown for comparison; the band defines the 1-$\sigma$ range of the
  mixing parameter estimate in \cite{bib:pdg2010}, which does
  not include the current result.}
\label{fig:pee}
\end{figure}

Our best value for the interaction rate of 862\,keV \ber\ solar
neutrinos in Borexino is 46.0$\pm$1.5$({\rm stat})$$^{+1.5}_{-1.6}({\rm
  syst})$\,\cpd.  If the neutrinos are assumed to be purely $\nu_e$, this
corresponds\footnote{In converting from interaction rates to fluxes, we use the electron scattering cross section from
\cite{bib:BahcallRadiativeCorrection}, with updated radiative correction
parameters from \cite{bib:pdg2010,bib:erlerRadCorr}, and a scintillator
electron density of (3.307$\pm$0.003)$\times$$10^{29}$/ton.} to an 862\,keV \ber\ solar neutrino flux\footnote{Note the
  distinction between the ``\ber\ solar neutrino flux'' and the ``862\,keV
  \ber\ solar neutrino flux'': the latter is an 89.6\% branch of the
  former.} of (2.78$\pm$0.13)$\times$$10^9$\,cm$^{-2}$s$^{-1}$. The corresponding flux
prediction from the SSM is
(4.48$\pm$0.31)$\times$$10^9$\,cm$^{-2}$s$^{-1}$, which, if all the
neutrinos remained $\nu_e$, would yield an interaction rate of
74.0$\pm$5.2\,\cpd\ in Borexino; the observed interaction rate is
5.0\,$\sigma$ lower. The ratio of
the measured to the predicted $\nu_e$-equivalent flux 
is 0.62$\pm$0.05. Under the assumption that
the reduction in the apparent flux is the result of $\nu_e$
oscillation to $\nu_\mu$ or $\nu_\tau$ (which undergo electron
elastic scattering interactions, but with a
cross section about 4.5 times lower than $\nu_e$ at this energy), we
find \Pee=0.51$\pm$0.07 at
862\,keV. The improved constraint on the low
energy solar \Pee\ is shown in Fig.~\ref{fig:pee}. 

We have also
performed a global analysis to determine the MSW neutrino mixing parameters in the two-flavour
approximation. This included the rate and spectrum information from the
other solar neutrino experiments
\cite{bib:rchem-cl,bib:rchem-ga,bib:kamiokande,bib:sno}, the SSM
flux predictions~\cite{bib:ssm2011} (with the exception that the \bor\ flux was
left free), and the current result.  We find best fit oscillation
parameters of $\tan^2{\theta_{12}}$=0.468$^{+0.039}_{-0.030}$ and $\Delta
m_{12}^2$=(5.2$^{+1.5}_{-0.9}$)$\times$$10^{-5}$\,eV$^2$; including
the KamLAND reactor anti-neutrino data~\cite{bib:kamland}
these become $\tan^2{\theta_{12}}$=0.457$^{+0.033}_{-0.025}$ and $\Delta
m_{12}^2$=(7.50$^{+0.16}_{-0.24}$)$\times$$10^{-5}$\,eV$^2$. The new result slightly
improves the precision with which the MSW mixing parameters can be determined.

Alternatively, by assuming MSW-LMA solar neutrino
oscillations, the Borexino results can be used to measure
the \ber\ solar neutrino flux. Using the oscillation parameters from
\cite{bib:pdg2010}, the Borexino result corresponds to a
total \ber\ neutrino flux
$\Phi_{\rm ^7Be}$=(4.84$\pm$0.24)$\times$10$^9$\,cm$^{-2}$s$^{-1}$ (after oscillation effects, the SSM prediction corresponds to an
862\,keV \ber\ neutrino interaction rate of 47.5$\pm$3.4\,\cpd\ in Borexino).
The ratio of our measurement to the SSM prediction gives \fbe=0.97$\pm$0.09.

Finally, we have used our new result to update the global
experimental constraint on the other low energy solar neutrino
fluxes by performing a global analysis in which the \bor, \ber, CNO, and
$pp$ fluxes were left as free parameters, following~\cite{bib:gonzalez}. Under the luminosity
constraint, we find $\Phi_{pp}$=(6.06$^{+0.02}_{-0.06}$)$\times$$10^{10}$\,cm$^{-2}$s$^{-1}$
and $\Phi_{{\rm CNO}}$$<$1.3$\times$$10^9$\,cm$^{-2}$s$^{-1}$~at
95\%~C.L. Expressed as a fraction of the SSM predicted fluxes
these correspond to $f_{pp}$=1.013$^{+0.003}_{-0.010}$ and
\fCNO$<$2.5~at 95\%~C.L. The latter limits the CNO
contribution to the solar luminosity to $<$1.7\%~(95\%~C.L.). Both the
precision of the {\it pp} flux determination and the constraint on the
CNO flux are improved by approximately a factor of two by our new result.

We have presented a measurement of the interaction rate of 862\,keV
\ber\ solar neutrinos in Borexino with a total uncertainty less
than 5\%. This precise measurement has also improved
our experimental understanding of the other low energy solar neutrinos. The Borexino measurements have allowed us to test
and validate the MSW-LMA model in the vacuum oscillation regime and at the lower
energy edge of the transition region with unprecedented precision.

The Borexino program is made possible by funding from INFN (Italy),
NSF (USA), BMBF, DFG, and MPG (Germany), NRC Kurchatov Institute
(Russia), and MNiSW (Poland).  We acknowledge the generous support of the Laboratori Nazionali del Gran Sasso (LNGS).  We thank J.~Erler, P.~Langacker, E.~Lisi, and G.~Ridolfi for helpful discussions.


\begin{thebibliography}{00}

\bibitem{bib:rchem-cl} B.T.~Cleveland et al., Ap. J. {\bf 496}, 505 (1998);
K.~Lande and P.~Wildenhain, Nucl. Phys. B (Proc. Suppl.) {\bf 118}, 49 (2003);
R.~Davis, Nobel Prize Lecture (2002).
%\bibitem{bib:rchem-ga} W.~Hampel et al. (GALLEX Collaboration),
%Phys. Lett. B {\bf 447}, 127 (1999);J.N.~Abdurashitov et al. (SAGE
%collaboration), Phys. Rev. Lett. {\bf 83}, 4686 (1999);M.~Altmann et
%al. (GNO Collaboration), Phys. Lett. B {\bf 616}, 174 (2005).
\bibitem{bib:rchem-ga} F.~Kaether et al., Phys. Lett. B {\bf 685}, 47
  (2010); W. Hampel et. al. (GALLEX Collaboration), Phys. Lett. B {\bf
  447}, 127 (1999); J.N.~Abdurashitov et al. (SAGE collaboration), Phys. Rev. C {\bf 80}, 015807 (2009).
\bibitem{bib:kamiokande} K.S.~Hirata et al. (KamiokaNDE
  Collaboration), Phys. Rev. Lett. {\bf 63}, 16 (1989);
Y. Fukuda et al. (Super-Kamiokande Collaboration), Phys. Rev. Lett. {\bf 81}, 1562 (1998);
J.P. Cravens et al. (SuperKamiokaNDE Collaboration), Phys. Rev. D {\bf
  78}, 032002 (2008).
\bibitem{bib:sno} Q.R.~Ahmad et al. (SNO Collaboration), Phys. Rev. Lett. {\bf 87}, 071301 (2001);
B.~Aharmim et al. (SNO Collaboration), Phys. Rev. C {\bf 75}, 045502 (2007);
B.~Aharmim et al. (SNO Collaboration), Phys. Rev. C {\bf 81}, 055504 (2010).
\bibitem{bib:bahcallSSM} J.N. Bahcall, A. Serenelli, and S. Basu,
  Ap. J. Suppl. {\bf 165}, 400 (2006).
\bibitem{bib:msw} S.P.~Mikheyev and A.Yu.~Smirnov, Sov. J. Nucl. Phys. {\bf 42}, 913 (1985);
L.~Wolfenstein, Phys. Rev. D {\bf 17}, 2369 (1978);
P.C.~de~Holanda and A.Yu.~Smirnov, JCAP {\bf 0302}, 001 (2003).
\bibitem{bib:kamland} S.~Abe et al. (KamLAND Collaboration),
  Phys. Rev. Lett. {\bf 100}, 221803 (2008).
\bibitem{bib:nonstandard} A. Friedland et al.,
  Phys. Lett. B {\bf 594}, 347 (2004); S. Davidson et al., JHEP
  {\bf 0303}, 011 (2003); P.C. de Holanda and A. Yu. Smirnov,
  Phys. Rev. D {\bf
    69}, 113002 (2004); A. Palazzo and J.W.F. Valle, Phys. Rev. D {\bf
    80}, 091301 (2009).
\bibitem{bib:bahcallLow}J.N. Bahcall in {\it Low Energy Solar Neutrino
  Detection}, Proceedings of the 2nd International Workshop,
  eds. Y. Susuki, M. Nakahata, and S. Moriyama (World Scientific,
  2002) pp. 172. \eprint{hep-ex/0106086}.
\bibitem{bib:bxdetectorpaper} G.~Alimonti et al. (Borexino Collaboration), Nucl. Instr. and Meth. A {\bf 600}, 58 (2009).
\bibitem{bib:bxfirstresults} C.~Arpesella et al. (Borexino Collaboration), Phys. Lett. B {\bf 568}, 101 (2008).
\bibitem{bib:bxsecondresults} C.~Arpesella et al. (Borexino
  Collaboration), Phys. Rev. Lett. {\bf 101}, 091302 (2008).
\bibitem{bib:ssm2011} A.M.~Serenelli, W.C.~Haxton and C.~Pe\~na-Garay,
  \eprint{arXiv:1104.1639}.
\bibitem{bib:pdg2010} Review of Particle Physics, K. Nakamura et al. (Particle Data Group), J. Phys. G {\bf 37}, 075021 (2010). 
\bibitem{bib:gatti} E.~Gatti et al., Energia Nucleare {\bf 17}, 34
  (1970). http://www-3.unipv.it/donati/papers/6d.pdf. 
%\bibitem{bib:osc} V.~Gribov and B.~Pontecorvo, Phys. Lett. B {\bf
%28}, 493 (1969); J.N.~Bahcall and R. Davis, Science {\bf 191}, 264
%(1976); J.N.~Bahcall and H.A.~Bethe, Phys. Rev. Lett. {\bf 65}, 2233 (1990).

\bibitem{bib:BahcallRadiativeCorrection} J.N. Bahcall,
  M. Kamionkowski and A. Sirlin, Phys. Rev. D {\bf 51}, 6146 (1995).
\bibitem{bib:erlerRadCorr}J. Erler and M.J. Ramsey-Musolf,
  Phys. Rev. D {\bf 72}, 073003 (2005).
%\bibitem{bib:uperKatmospheric} Y.~Ashie et al.,(SuperKamiokaNDE Collaboration) Phys. Rev. Lett. {\bf 93}, 101801 (2004).
%\bibitem{bib:k2k} M.H.~Ahn et al. (K2K Collaboration), Phys. Rev. D 74, 072003 (2006).
%\bibitem{bib:minos} P.~Adamson et al. (MINOS Collaboration), Phys. Rev. Lett. {\bf 101}, 131802 (2008).
%\bibitem{bib:opera} N. Agafonova et al., OPERA Collaboration, Phys.Lett. B {\bf 691}, 138 (2010).
%\bibitem{bib:bxtechnology} G.~Alimonti et al. (Borexino Collaboration), Astropart.  Phys. {\bf 16}, 205 (2002).
%\bibitem{bib:bxvessels} J.~Benziger et al., Nucl. Inst. and Meth. A {\bf 582}, 509 (2007).
%\bibitem{bib:bxdmp} M.~Chen et al., Nucl. Inst. and Meth. A {\bf 189}, (1999).
%\bibitem{bib:bxpmts} A.~Ianni et al., Nucl. Inst. and Meth. A {\bf
%537}, 683 (2005);
%A.~Brigatti et al., Nucl. Inst. and Meth. A {\bf 537}, 521 (2005).
%\bibitem{bib:bxcones} L.~Oberauer et al., Nucl. Inst. and Meth. A {\bf 530}, 453 (2004).
%\bibitem{bib:bxmuondetector} G.~Bellini et al. (Borexino Collaboration), \eprint{arXiv:1101.3101v1}, accepted for publication in JINST.
%\bibitem{bib:bxrad} C.~Arpesella et al. (Borexino Collaboration), Astropart. Phys. {\bf 18}, 1 (2002).
%\bibitem{bib:bxpur} J.~Benziger et al.,  Nucl. Inst. and Meth. A {\bf 587}, 277 (2008).
%\bibitem{bib:bxlakn} H.~Simgen and G.~Zuzel, {\it Ultrapure gases - From the Production Plant to the Laboratory}, AIP Conference Proceedings Vol.~897, Topical Workshop on Low Radioactivity Techniques: LRT 2006, Aussois (France), pp. 45--50, ed. P.~Loaiza, Springer (2007).
%\bibitem{bib:bx-be7long} G.~Bellini et al. (Borexino Collaboration), detailed paper on \ber\ measurement, in preparation.
%\bibitem{bib:bxc14} G.~Alimonti et al. (Borexino Collaboration), Phys. Lett. B {\bf 422}, 349 (1998).
%\bibitem{bib:bxab} H.O.~Back et al. (Borexino Collaboration), Nucl Inst. Meth. {\bf 584}, 98 (2008).
%\bibitem{bib:birks}J.B.~Birks, Proc. Phys. Soc. A, {\bf 64}, 874 (1951).
%\bibitem{bib:bxcalib} G.~Bellini et al. (Borexino Collaboration), paper on calibrations, in preparation.
%\bibitem{bib:bxboron} G.~Bellini et al. (Borexino Collaboration), Phys. Rev. D {\bf 82}, 033006 (2010).
\bibitem{bib:barger} V.~Barger, D.~Marfatia and K.~Whisnant,
  Phys. Lett. B {\bf 617}, 78 (2005). 
%\bibitem{bib:bxboron} G.~Bellini et al. (Borexino Collaboration), Phys. Rev. D {\bf 82}, 033006 (2010).
\bibitem{bib:gonzalez} M.C.~Gonzalez-Garcia, M.~Maltoni and J.~Salvado, JHEP {\bf 072}, 1005 (2010).

\end{thebibliography}
\end{document}